\documentclass[12pt]{emulateapj}

\usepackage{graphics}

\newcommand{\omc}{\hbox{$\omega$ Cen~}}

\newcommand{\omcen}{\hbox{$\omega$ Centauri~}}

\shorttitle{More than two thousand WDs in {\boldmath
$\omega$}\,Cen} \shortauthors{Monelli et al.}

\begin{document}
\title{Discovery of more than two thousand white dwarfs in the globular cluster {\boldmath $\omega$}\,Centauri\altaffilmark{1}}
\author{
M.~Monelli \altaffilmark{2},
C.E.~Corsi \altaffilmark{2},
V.~Castellani \altaffilmark{2,3}, 
I.~Ferraro \altaffilmark{2},
G.~Iannicola \altaffilmark{2},
P.G.~Prada Moroni \altaffilmark{4,5,6},
G.~Bono \altaffilmark{2},
R.~Buonanno \altaffilmark{7},
A.~Calamida \altaffilmark{2,7},
L.M.~Freyhammer \altaffilmark{8,9}, 
L.~Pulone \altaffilmark{2}, and 
P.B.~Stetson \altaffilmark{10}
}

\altaffiltext{1}
   {Based on observations collected with the Advanced Camera for Surveys on
    board of the Hubble Space Telescope.}
  \altaffiltext{2}
    {INAF-Osservatorio Astronomico di Roma, via Frascati 33,
     Monte Porzio Catone, Rome, Italy; bono@mporzio.astro.it, corsi@mporzio.astro.it, 
     ferraro@mporzio.astro.it, giacinto@mporzio.astro.it, monelli@mporzio.astro.it, 
     pulone@mporzio.astro.it}
  \altaffiltext{3}
     {INFN, Sezione di Ferrara, via Paradiso 12, 44100 Ferrara, Italy} 
  \altaffiltext{4}
   {Dipartimento di Fisica "E. Fermi", Univ. Pisa, Largo B. Pontecorvo 2, 56127 Pisa, 
    Italy; prada@df.unipi.it}
  \altaffiltext{5}
    {INFN, Sez. Pisa, via E. Fermi 2, 56127 Pisa, Italy}
  \altaffiltext{6}
    {INAF-Osservatorio Astronomico di Teramo, via M. Maggini, 64100 Teramo, Italy}
  \altaffiltext{7}
    {Universit\`a di Roma Tor Vergata, via della Ricerca Scientifica 1, 00133 Rome,
     Italy; buonanno@mporzio.astro.it, calamida@mporzio.astro.it}
  \altaffiltext{8}
    {Royal Observatory of Belgium, Ringlaan 3, B-1180 Brussels, Belgium}
  \altaffiltext{9}
    {Vrije Universiteit Brussel, OBSS/WE, Pleinlaan 2, B-1050 Brussels, Belgium;
   lfreyham@vub.ac.be}
  \altaffiltext{10}
     {Dominion Astrophysical Observatory, Herzberg Institute of
      Astrophysics, National Research Council, 5071 West Saanich Road, Victoria,
      British Columbia V9E 2E7, Canada; Peter.Stetson@nrc-cnrc.gc.ca}

%
\date{\centering drafted \today\ / Received / Accepted }

\begin{abstract}
We present deep multiband (F435W, F625W, and F658N) photometric
data of the Globular Cluster \omc collected with the Advanced
Camera for Surveys on board of the Hubble Space Telescope. We
identified in the (F435W-F625W, F435W) plane more than two
thousand White Dwarf (WD) candidates using three out of nine
available pointings. Such a large sample appears in agreement with 
predictions based on the ratio between WD and Horizontal Branch 
(HB) evolutionary lifetimes. We also detected  $\approx 1600$ WDs 
in the (F658N-F625W, F625W) plane, supporting the evidence that a 
large fraction of current cluster WDs are $H_\alpha$ bright.    
\end{abstract}

\keywords{globular clusters: general --- globular clusters: omega Centauri }

\maketitle

\section{Introduction}

Among Galactic Globular Clusters (GGCs), \omc is the most massive
one ($M=5\times10^6\, M_\odot$, Meylan et al. 1995) and the only
one which clearly shows a well-defined spread in metallicity.
According to recent estimates based on sizable samples of evolved
Red Giant Branch (RGB) and sub-giant branch stars, the metallicity 
distribution shows three peaks around ${[\rm Fe/H}]=-1.7$, $-1.5$ 
and $-1.2$ together with a tail of metal-rich stars approaching ${[\rm
Fe/H}]\approx-0.5$ (Norris et al.\ 1996; Hilker et al.\ 2004). 
During the last few years it has also been
suggested that \omc harbors multiple stellar populations (Lee et
al.\ 1999) characterized by different ages (Ferraro et al.\ 2004;
Hughes et al.\ 2004), He abundances, and distances (Bedin et
al.\ 2004; Norris 2004; Freyhammer et al.\ 2005). 
More recently, Sollima et al. (2005) suggested that RGB stars
present five peaks due to different stellar populations. 
Moreover, Piotto et al. (2005) found, that the bluer main
sequence (MS) detected by Bedin et al. (2004) is more metal-rich than
the red MS. To account for this feature they suggested a possible
variation of the He content among MS stars.

Even though \omc presents properties which need to be properly
understood, its stellar content is a gold mine to investigate
several open problems concerning stellar evolution and its
dependence on the metallicity. This applies not only to evolved
stars such as RR Lyrae, hot HB stars, and the tip of the RGB, 
but also to the expected population(s) of WDs. The
search for WDs in GGCs has already been successful, and several WD
samples have been identified (Hansen et al.\ 2003; Moehler et al.\
2004). So far, the richest sample of
cluster WDs (222) was detected in M4 by Hansen et al. (2004). The
detection of WDs in \omc dates back to  Ortolani \& Rosino (1987)
who selected two dozen WD candidates on the basis of ground-based
data, and to Elson et al. (1995), who detected four WD candidates
using HST data. In this paper, we present
preliminary results concerning the identification of more than
2,200 WDs in \omc on the basis of $B,R,H_\alpha$ data collected
with the Advanced Camera for Surveys (ACS) on board of HST, and
available on the HST archive.
The reader interested in thorough reviews concerning recent
theoretical and empirical results is referred to (Fontaine,
Brassard, \& Bergeron 2001; Koester 2002;  Prada Moroni \& 
Straniero 2002; Hansen \& Liebert 2003; Hansen 2004, and 
references therein).

\section{Observations and data reduction}

Current data were collected with nine pointings of the ACS camera
across the center of the cluster. The $3\times3$ mosaic covers a
field of view of $\approx9\arcmin \times9\arcmin$. Four images per
field have been acquired in three different bands, namely F435W,
F625W, and F658N (hereinafter $B$, $R$, and $H_\alpha$). Three
deep (340s) exposures were secured for the B and R-band,
respectively, while the exposure time for the four $H_\alpha$
images was 440s each. The nine fields were
independently reduced with the DAOPHOTII/ALLFRAME (Stetson 1994).
An individual PSF has been extracted for each frame by adopting,
on average, $\approx$200 bright isolated stars. The individual
catalogues were rescaled to a common geometrical system with
DAOMATCH/DAOMASTER. The final catalogue includes approximately
1.2$\times 10^6$ stars, and the $B,B-R$ CMD shows a well-populated
cluster MS together with a sizable sample of stars covering a wide
region to the left of the MS (Bono et al. 2005).

From this catalogue we selected all the stars bluer than MS stars 
and fainter than extreme HB stars. We ended up with
a sample of approximately 45,000 stars distributed over the nine 
pointings. Among them we selected three pointings (2, 7, and 9)
that include $\approx 14,000$ WD candidates. These stars have been
identified in individual deep $B,R,H_\alpha$ images and we performed 
once again the photometry using ROMAFOTwo. Individual stars have been 
interactively checked in every image, 
and the WD candidates were 
measured either as isolated stars or together with neighbor stars, 
where crowding affects the photometric accuracy. Note that a substantial 
fraction of the stars located close to WD candidates are truly MS stars, 
i.e. they did not belong to the original sample of stars located on the 
blue side of the MS. 
Note that the 80\% of the selected stars revealed to be either 
cosmic rays, or spurious identifications of faint stars located close 
to saturated stars, or too faint objects to be safely measured on
individual images. 
The photometric calibration was performed in the Vega System
(http://www.stsci.edu/hst/acs/documents).

\section{Results and Discussion}

\noindent 
In this investigation we present preliminary results based on three 
out of the nine pointings. Figure 1 shows the 
current sample of
\omcen stars in the $B-R$, $B$ plane. Interestingly enough, the
original sample splits into two well-defined sequences, the redder
one made by MS stars and the other one made by 2,212 blue objects
that ranges from $B\approx 22$, $B-R\approx 0$, down to $B\approx
27$, $B-R\approx 0.8$, covering the expected region of cluster WDs.

Using GOODS data collected with ACS in the bands F435W and F606W,
we estimated that the expected number of field galaxies with
$22\le B \le 26$  and $\-0.2 \le B-R \le 0.5$\footnote{Note that
to perform this estimate we accounted for the difference in
magnitude between AB and Vega systems and for the difference
between the filter F606W and the filter F625W
(http://www.stsci.edu/hst/acs/documents).} in the same area
covered by current observations is $\approx 60$ (Grazian et al.
2005, priv. comm.). The number of field stars is also
negligible, because  halo and disk stars peak around $B-R=0.7$ and
$B-R=1.8$, respectively (King et al. 1990). More detailed
estimates based on radial velocity (Suntzeff \& Kraft 1996) or
proper motion (van Leeuwen et al. 2000) measurements in \omc 
suggest that at most two dozen of field stars might be located 
inside the area covered by current data. Finally, theoretical 
Galactic models (Castellani et al. 2002) suggest that in the 
same area covered by current observations are present at most 
sixteen field WDs brighter than $B=25.5$ and fifty-five field 
WDs brighter than $B=28$. 
%

This evidence suggests that we are facing with a bona fide sample
of cluster WDs, including more than 2,000 objects, thus the
largest sample of WDs ever observed in a stellar cluster. Data 
plotted in Fig. 1 clearly show that, thanks to the sizable sample 
of \omc stars, the cooling sequence show up at $B\approx 21$. This 
bright limit, if we assume for \omc an apparent distance modulus 
$DM_B\sim 14.21$ (Thompson et al. 2001), implies that current data 
are tracing the cooling history of cluster WDs at least below $M_B\sim 7$.
According to predictions by Althaus, \& Benvenuto (1998) for a $0.5 M_\odot$ WD 
with a CO core and pure H atmosphere models by Bergeron, Wesemael, \& Beauchamp 
(1995)\footnote{See also http://www.ASTRO.UMontreal.CA/~bergeron/CoolingModels/} 
the current WD sample provides the opportunity to investigate the WD cooling 
for luminosities ranging from ten times the solar luminosity 
down to $\log L/L_\odot \sim -3.1$ ($B \sim 27$, $M_B \sim 12.8$).

By adopting the same theoretical predictions one can easily recognize 
that the huge number of WDs should not be a real surprise. As a matter 
of fact, the number ratio between WD and HB stars, for not too long 
cooling times, is simply given by the ratio of the lifetimes spent during  
these two evolutionary phases. On the basis of the shallow ACS photometry 
in the $B,R$ bands, recently presented by Freyhammer et al. (2005), we found 
in the same three fields the occurrence of $\sim 630$ HB stars. The typical 
evolutionary lifetime for HB stars is $\sim 1-2\times10^8$ year, depending 
on their ZAHB effective temperature (Castellani et al. 2004), while for a 
WD with a CO core and $M=0.5 M_\odot$ at $M_B \sim 11.3$ ($\log L/L_\odot= -2.24$)  
the lifetime is $\sim 3.5\times10^8$ years. This means that we expect to 
detect roughly twice as many WDs than HB stars brighter than $B\approx25.5$.
Interestingly enough, we detected approximately 1200 WDs brighter than this 
magnitude limit. However, the current estimate should be considered as a robust 
lower limit, since preliminary artificial star experiments suggest that 
at this limiting magnitude the completeness is $\approx 85$\%.

An unexpected observed feature in Fig. 1, is the steady increase in 
color dispersion when moving toward fainter WD magnitudes. In order
to assess whether this spread in color might be due to photometric
errors, we performed an empirical test. We estimated the ridge
line of WDs and the distance in color of individual objects from
the ridge line. Figure 2 shows the color distribution of WDs in
the magnitude interval $B=25\pm0.5$ together with its gaussian fit
(solid line). Then we estimated standard deviation in color of the
same selected WDs (intrinsic errors) and we found $\sigma_{B-R}=0.087$. 
The dashed line plotted in Fig. 2 shows the expected color distribution 
for the same sample of WDs (712) according to the assumption that their 
$B,R$ magnitudes would only be affected by gaussian intrinsic errors. 
Data plotted in Fig. 2 indicate that the sigma of the gaussian fit to 
observed WDs is a factor of two larger than expected. On this basis, 
we estimated that the two distributions differ at 99\% confidence level. 
This finding, taken at face value, indicates that the color dispersion 
might be real.


In order to investigate the WD location in the CM diagram we adopted the
WD cooling sequences for CO core and H envelopes constructed by
Althaus, \& Benvenuto (1998). Theoretical predictions have been
transformed into the observational plane by using pure H atmosphere 
models constructed by Bergeron et al. (1995).
Predicted cooling sequences for $M=0.5, 0.7, 0.9 M_\odot$ have
been plotted by adopting canonical estimates for cluster reddening
($E(B-V)=0.11\pm0.02$, Lub 2002) and distance modulus
($\mu=13.7\pm 0.11$, Thompson et al. 2001). The reddening in the
$B,R,H_\alpha$ bands was estimated using the extinction model of
Cardelli et al. (1989).

The comparison between theory and observations (see panel a) of 
Fig. 3) discloses a  good
agreement for the bright portion ($22 \le B \le 23$). However,
the theoretical sequences toward fainter magnitudes appear to fit  
the blue (hot) edge of the observed WDs. It is noteworthy, that  
a WD with a stellar mass $M=0.5 M_\odot$ is a lower limit for 
WDs with CO-core, actual WDs are expected to be slightly more 
massive ($\sim 0.53 M_\odot$, Renzini et al. 1996). Current
uncertainties on reddening (0.02 across the cluster,
Schlegel et al. 1998) and on cluster distance cannot account for 
the observed systematic drift in color. This evidence suggests that
the observed WD cooling sequence is redder than expected.

In principle one can find several plausible reasons for such an 
occurrence. In particular, we note that for $22 \le B$ we are 
already below the so-called "DB gap" (Hansen \& Liebert 2003), 
i.e. WDs with He atmospheres could be present. Data plotted in 
panel b) of Fig. 3, show that by adopting WD cooling sequences 
for CO core, and He envelopes by Benvenuto, \& Althaus (1997) 
together with He atmosphere models by Bergeron et al. (1995), 
the cooling sequence is indeed moving toward redder colors. 
Note that the cut-off of the cooling sequences in the bright 
region is due to the fact that He atmosphere models do not cover 
this temperature region. The comparison between theory and 
observations indicates that the bulk of the observed WDs might be 
of the DB type.  
This evidence, as suggested by the referee, is in contrast with current
spectroscopic findings by Moehler et al. (2004) concerning cluster WDs.
They identified of H Balmer lines in four and seven
WDs in NGC 6397 and in NGC 6752, thus suggesting that these objects are
of the DA type. The same outcome applies to field WDs for which DA WDs
are $\approx 80$\% of the entire sample (Koester \& Chanmugam 1990).
An explanation of this discrepancy should await for new spectroscopic 
measurements.

As shown in the bottom panel of Fig. 3, a different possibility is 
given by the occurrence of He-core WDs. Interestingly enough, 
predicted WD cooling sequences for He core structures by 
Serenelli et al. (2002), transformed into the observational plane 
by adopting H atmosphere models, also account for the observed 
distribution. 
This working hypothesis, once confirmed, indicates that a 
substantial fraction of WDs in \omc might be the evolutionary aftermath 
of binary systems. According to current knowledge, candidate cluster 
He core WDs have only been identified in systems with high central 
densities and short two body relaxation times (NGC6397,
$\log \rho_0 =5.68 \; L_\odot pc^{-3}$, Taylor et al. 2001;  
47 Tuc, $\log \rho_0 =4.77 \; L_\odot pc^{-3}$, Edmonds et al. 2001). 
However, \omc presents a relatively low central density
($\log \rho_0 =3.12 \; L_\odot pc^{-3}$), and therefore fills an empty
region of the parameter space predicted by Hansen et al. (2003, see
their figure 8). 
Current findings are marginally affected by the adopted 
theoretical predictions. Different sets of WD cooling sequences 
(Fontaine et al. 2001; Prada Moroni \& Straniero 2002),  
transformed into the $B-R, B$ plane by adopting the same atmosphere 
models, agree quite well with each other.  

Finally, it is worth noting that current $H_\alpha-R, R$ data 
(see Fig. 4), show that a good fraction of detected WDs ($\sim$ 1,600)
are $H_\alpha$ bright, and indeed they attain $H_\alpha-R$ colors 
systematically bluer than predicted. 
We would like to mention that blending and/or binarity 
could also affect observed colors. Therefore, no firm conclusion 
can be reached on the basis of current data. More detailed photometric 
and spectroscopic investigations of this large sample of cluster 
WDs will provide fundamental hints concerning their cooling  
and progenitors.

\section{Acknowledgments}
It is a real pleasure to thank P. Bergeron not only for sending us 
bolometric corrections and color indices for ACS bands, but also for 
many useful suggestions. We also thank M. Cignoni 
and S. Degl'Innocenti for predictions concerning Galactic models and 
isochrones. It is also with unusual pleasure that we acknowledge an 
anonymous referee for his/her positive comments and insights. 
This work was supported by PRIN~2003 and by BFSP within the projects: 
``Continuity and Discontinuity in the Galaxy Formation`` and ``IAP P5/36''.


\clearpage

\begin{figure}
\resizebox{1.08\hsize}{!}{\includegraphics*[viewport=1 -1 500 559,clip]{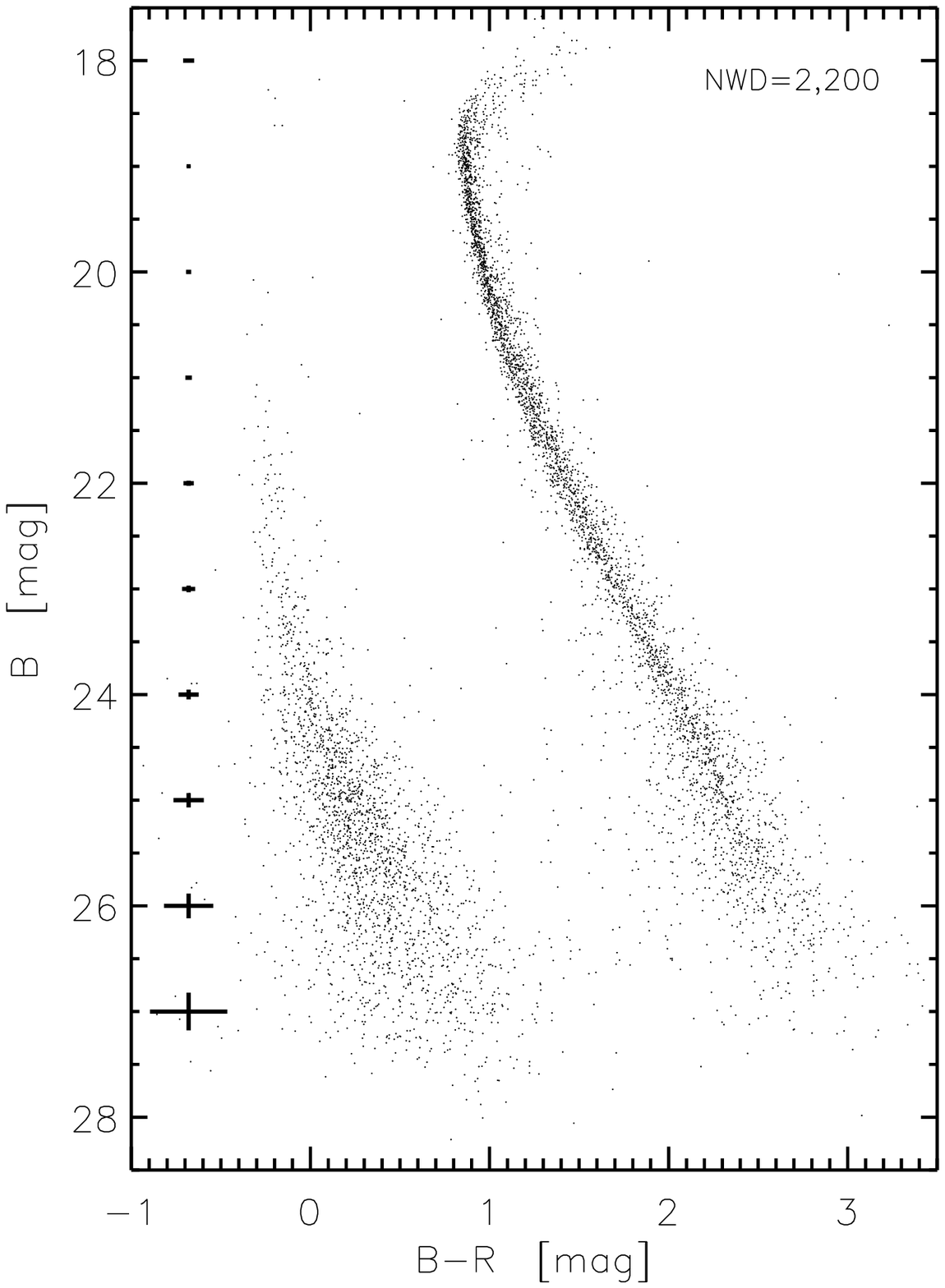}} 
\caption[]{Color-Magnitude diagram in the $B, B-R$ bands. The number of WD 
candidates are labeled. Error bars display intrinsic photometric errors.} 
\label{fig:apjfig1} 
\end{figure}

\clearpage

\begin{figure}
\resizebox{1.20\hsize}{!}{\includegraphics*[viewport=1 -1 490 500,clip]{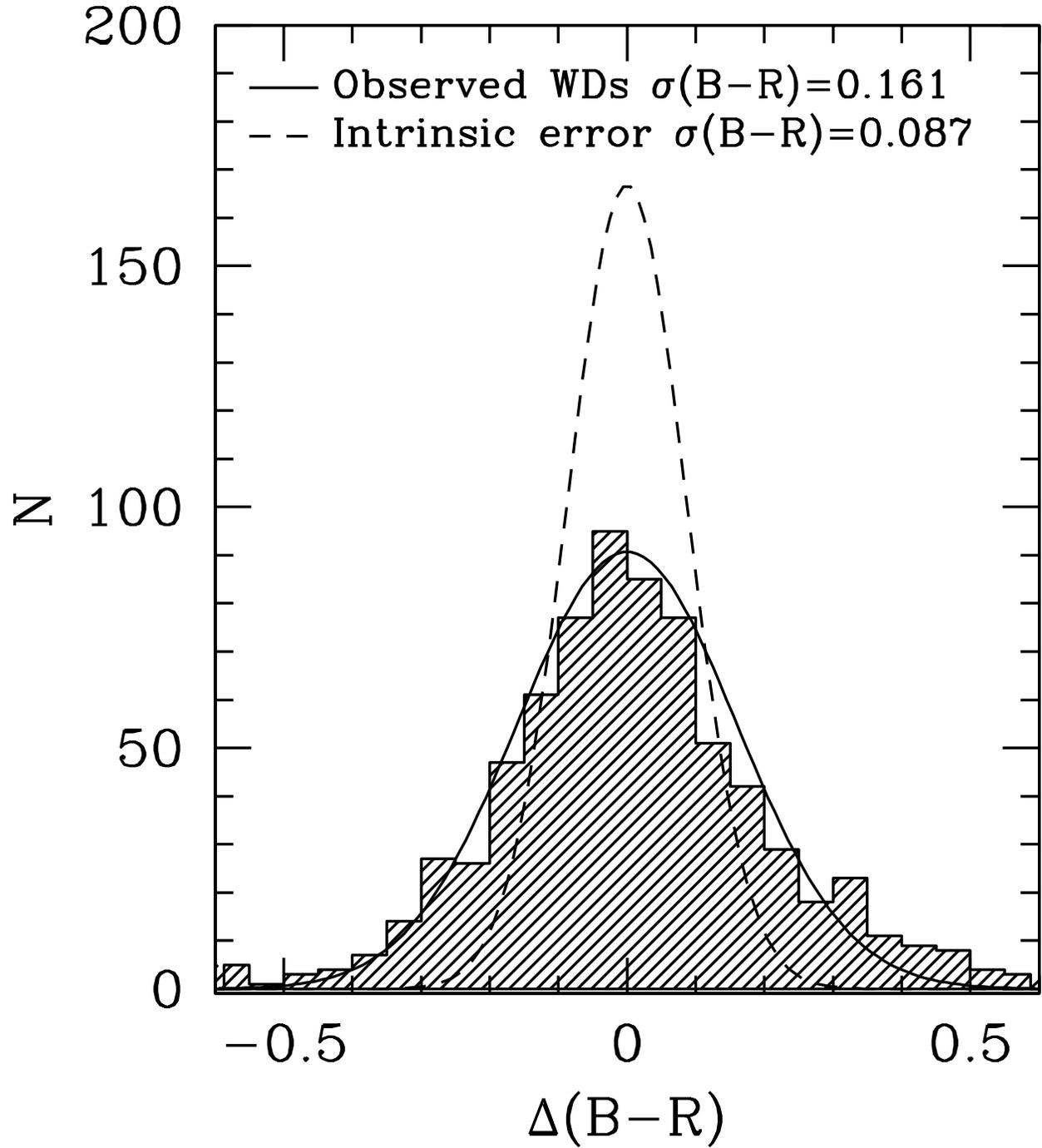}} 
\caption[]{Color distribution of WDs
in the magnitude interval $B=25\pm0.5$. The solid line shows the
gaussian fit to the observed distribution, while the dashed line
the expected distribution for the same sample of WDs in the case that their 
colors would only be affected by gaussian photometric
intrinsic errors. } \label{fig:apjfig2}
\end{figure}

\clearpage

\begin{figure}
\epsscale{.60}
\plotone{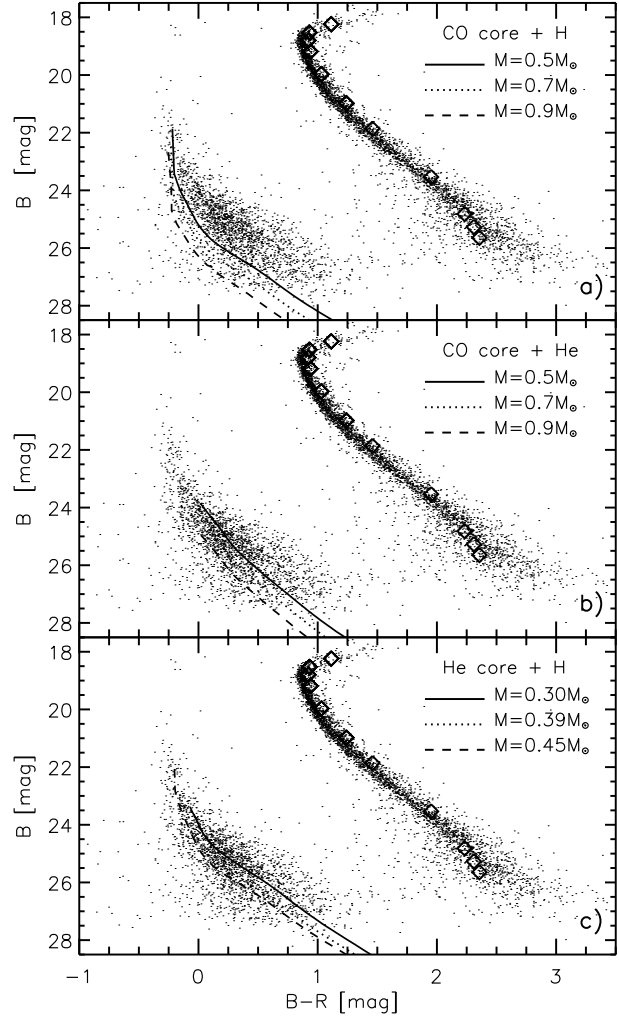}
\vspace{7mm}
\caption[]{Color-Magnitude diagram
in the $B, B-R$ bands. Panel {\bf a)} - Solid, dotted, and dashed
lines show predicted WD cooling sequences for CO core and H
envelopes with stellar masses equal to 0.5, 0.7, and 0.9
$M/M_\odot$ (Althaus, \& Benvenuto 1998). Theoretical predictions
have been transformed into the observational plane by adopting H
atmosphere models. Diamonds show a 12 Gyr isochrone for 
Z=0.001 and Y=0.232 constructed by Cariulo et al. (2004, $M\ge 0.5 M_\odot$) 
and by Baraffe et al. (1997, $M< 0.5 M_\odot$). Panel {\bf b)} - Same as panel 
{\bf a)}, but for WD cooling sequences, with CO core and He envelopes
(Benvenuto, \& Althaus 1997), transformed into the observational 
plane by adopting He atmosphere models.
Panel {\bf c)} - Same as panel {\bf a)}, but for WD cooling
sequences with He core and H envelopes (Serenelli et al. 2002),  
stellar masses equal to 0.30, 0.39, and 0.45 $M/M_\odot$
and H atmospheres.} \label{fig:apjfig3}
\end{figure}

\clearpage

\begin{figure}
\resizebox{0.88\hsize}{!}{\includegraphics*[viewport=10 -1 430 559,clip]{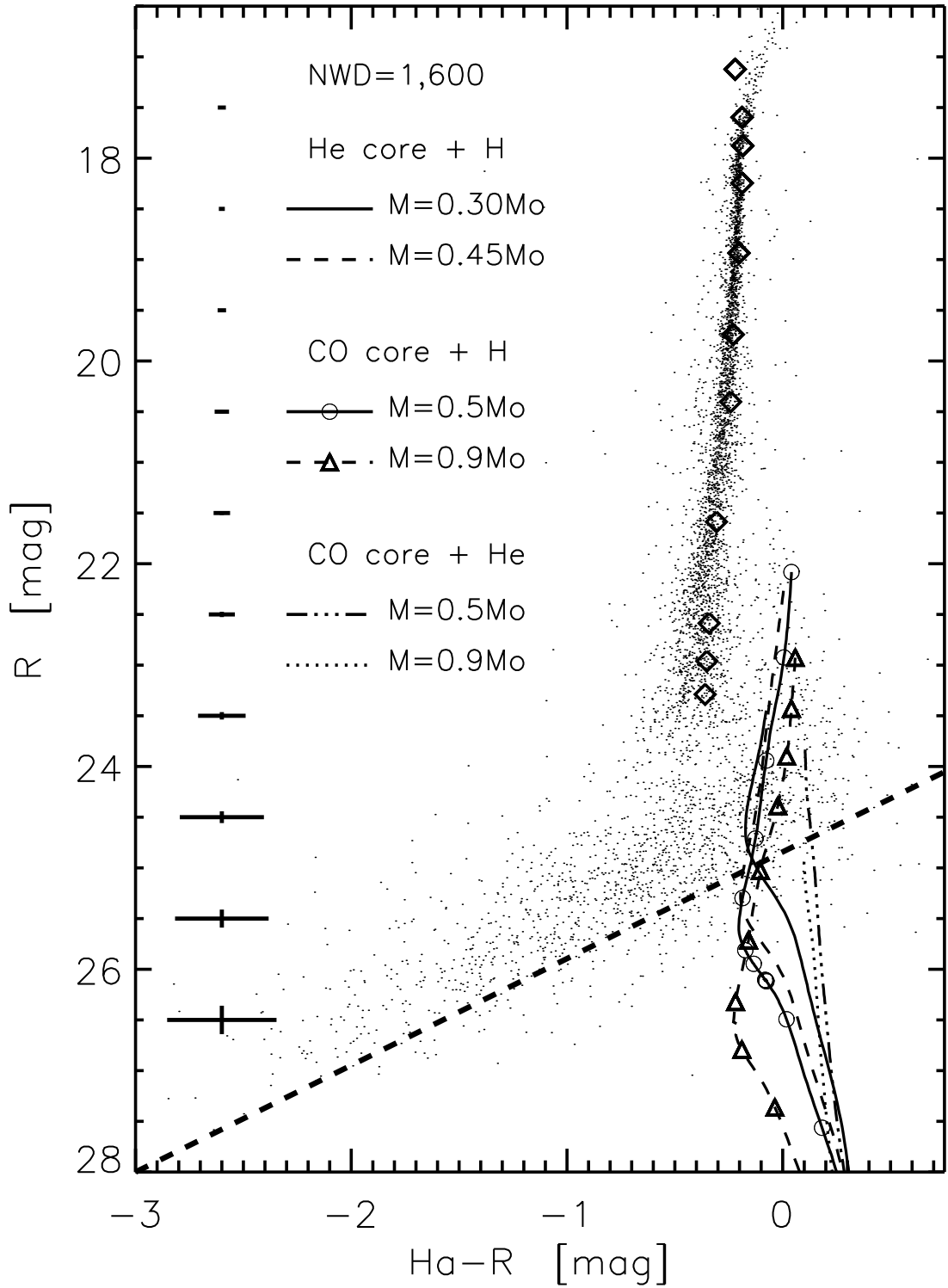}} 
\caption[]{Color-Magnitude diagram in the $R, H_\alpha-R$ bands. Predicted cooling 
sequences are the same as in Fig. 4. The thick dashed line marks the 
detection limit. Error bars display intrinsic photometric errors.}
\label{fig:apjfig4}
\end{figure}

\end{document}